\documentclass[final,5p,times,twocolumn]{elsarticle}

\usepackage[normalem]{ulem}
\usepackage{natbib}
\usepackage{subcaption}
\usepackage{amssymb}
\usepackage{amsthm}
\usepackage{amsmath}
\usepackage{amsfonts}
\usepackage{bbm}
\usepackage{mathrsfs}
\usepackage{multirow}
\usepackage{hyperref}
\usepackage{algorithm}
\usepackage{algpseudocode}
\usepackage{ctable}
\newcolumntype{K}{>{\centering\arraybackslash}X}

\newcommand{\vect}[1]{\mathbf{#1}}
\newcommand{\vers}[1]{\hat{\mathbf{#1}}}

\begin{document}

\begin{frontmatter}

\title{Diffusion of Globular Macromolecules in Liquid Crystals of Colloidal Cuboids}

\author[lab1]{Luca Tonti}
\author[lab1]{Fabián A. García Daza}
\author[lab1]{Alessandro Patti\corref{*}}
\cortext[*]{E-mail: alessandro.patti@manchester.ac.uk}
\address[lab1]{Department of Chemical Engineering and Analytical Science, The University of Manchester, Manchester, M13 9PL, UK}


\begin{abstract}
Macromolecular diffusion in dense colloidal suspensions is an intriguing topic of interdisciplinary relevance in Science and Engineering. While significant efforts have been undertaken to establish the impact of crowding on the dynamics of macromolecules, less clear is the role played by long-range ordering. In this work, we perform Dynamic Monte Carlo simulations to assess the importance of ordered crowding on the diffusion of globular macromolecules, here modelled as spherical tracers, in suspensions of colloidal cuboids. We first investigate the diffusion of such guest tracers in very weakly ordered host phases of cuboids and, by increasing density above the isotropic-to-nematic phase boundary, study the influence of long-range orientational ordering imposed by the occurrence of liquid-crystalline phases. To this end, we analyse a spectrum of dynamical properties that clarify the existence of slow and fast tracers and the extent of deviations from Gaussian behaviour. Our results unveil the existence of randomly oriented clusters of cuboids that display a relatively large size in dense isotropic phases, but are basically absent in the nematic phase. We believe that these clusters are responsible for a pronounced non-Gaussian dynamics that is much weaker in the nematic phase, where orientational ordering smooths out such structural heterogeneities.  

\end{abstract}

\begin{keyword}
Colloids \sep Brownian Motion \sep Diffusion \sep Liquid crystal \sep Dynamic Monte Carlo simulation
\end{keyword}



\end{frontmatter}


\section{Introduction}
\label{intro}

Understanding the diffusion of tracers in crowded colloidal suspensions is a problem with a multifaceted interdisciplinary impact. In Biology, it is especially related to the ability of macromolecules, such as proteins, of penetrating the cell membrane, diffusing through the cytosol and organelles and thus contributing to regulate the cell function \cite{dix2008}. In Nanomedicine, the release rate of a drug from a nanovehicle, such as a micelle, depends on its ability to diffuse through its hydrophobic core and hydrophilic corona \cite{kamaly2016}. In Food Science, moisture migration in dried food products determines how long these can be preserved from spoilage \cite{biji2015}. In Materials Science, the self-healing ability of a smart coating, which is activated by an external stimulus, such as a pH or temperature gradient, relies on the diffusion of a corrosion inhibitor through a polymeric matrix into the defect \cite{wei2015}. Additionally, investigating the dynamics of tracers in colloids has also opened a path to the study of rheology of the host phase, a technique commonly referred to as microrheology, which allows to extract the viscoelastic response of a soft material from a tracer's mean square displacement (MSD) \cite{puertas2014,Habibi2019}.

Although specific attributes and properties make them unique and different from each other, all the above-mentioned soft-matter systems exhibit interesting common features. One is that macromolecular diffusion shows an \textit{anomalous} behaviour,
where the MSD, rather than growing linearly with time, as predicted by Fick's diffusion theory \cite{FICK1995}, follows a power-law of the type $r^2 (t) \propto t^\beta$, with $0<\beta<1$ the anomalous diffusion exponent \cite{BOUCHAUD1990, METZLER2000, metzler2014, Hanneken2004}. Such a subdiffusion usually develops at intermediate time scales, but it is not persistent and, at sufficiently long times, the MSD recovers a Brownian-like behaviour, generally referred to as Fickian diffusion \cite{sokolov2012, ghosh2015, cuetos2018}. Crowded soft-matter systems can also exhibit a degree of ordering that might influence the diffusion of macromolecules or nanoparticles, for instance by creating preferential paths \cite{burgos2017}, and thus should not \textit{a priori} be disregarded. Nevertheless, when assessing  macromolecular diffusion in crowded media, most research works have especially focused on the host-guest affinity \cite{groen2015, nogueira2020}, relative characteristic lengths \cite{gam2011, dey2019} and volume fractions \cite{Roosen-Runge2011, Chakrabarti2013} as well as on the effect of structural heterogeneities \cite{smith2017}, but much less on whether or not it could be enhanced, inhibited or made more complex by the occurrence of long-range ordering and spatial anisotropy.

Due to their rich phase behaviour, colloidal liquid crystals (LCs), namely colloidal suspensions of orientationally and/or positionally ordered anisotropic particles, are particularly appropriate to gain insight into the effect of ordered crowding on macromolecular diffusion. This is especially evident in systems of biaxial particles, whose phase behaviour is decorated by a plethora of intriguing morphologies that cannot be observed with uniaxial particles \cite{glotzer2007}. Over the last few years, our group has specifically explored the self-assembly of oblate and prolate cuboidal particles, unveiling the existence of uniaxial and biaxial nematic and positionally-ordered smectic and columnar phases \cite{Cuetos2017, patti2018, cuetos2019, effran2020}. More recently, we have also determined the main features of their equilibrium dynamics in uniaxial nematics, detecting Fickian and Gaussian dynamics at both short-time and long-time scales 
\cite{cuetos2020}. In this work, we investigate the diffusion of globular macromolecules, here modelled as spherical tracers, in uniaxial nematics of oblate and prolate colloidal cuboids, modelled as hard board-like particles (HBPs). Our main goal is understanding how long-range ordering affects tracers' diffusion as compared to diffusion in weakly ordered (isotropic) phases. To this end, we employ a stochastic simulation technique, referred to as Dynamic Monte Carlo (DMC), that is especially efficient to mimic the Brownian motion of particles interacting by mere excluded volume effects \cite{Patti2012, Cuetos2015, Corbett2018, Daza2020, chiappini2020}. 


\section{Model and simulation methodology}
\label{method}

The systems studied in this work comprise $N_{c} = 2000$ HBPs of thickness $T$, which is the system unit length, reduced length $L^* \equiv L / T$ and width $W^* \equiv W / T$, and $N_{s} = 200$ hard spheres (HSs) with diameter ${d_{s}}^* \equiv d_{s} / T$. The size of the spherical tracers with respect to that of cuboids has been set to mimic the diffusion of macromolecules, such as globular proteins like enzymes, whose diameter is generally in the order of 1 to 10 nm \cite{Erickson2009}. Given that these model macromolecular tracers are incorporated in a host phase of significantly larger colloidal particles, their diameter has been set to $d_{S} = T / 10$. In particular, we investigated the dynamics of spherical tracers in systems of HBPs with reduced dimensions $(L^*,W^*) = \{(12,1), (12,8)\}$ at packing fractions $\eta = 0.15$ and $\eta = 0.34$, where HBPs self-assemble, respectively, into isotropic ($\rm I$) and uniaxial nematic ($\rm N_U$) phases \cite{Cuetos2017}. The packing fraction is defined as $\eta^{pure} \equiv LWTN_c/V$, with $V$ the volume of the simulation box.
Initial configurations have been prepared by incorporating the tracers into pure systems of cuboids at the desired packing fraction $\eta^{pure}$. The actual packing fraction of the mixture should also include the volume occupied by the spherical tracers as follows 
\begin{gather}
    \eta = \eta^{pure} \left[ 1 + \delta \eta\right], \\
    \delta \eta = \frac{\pi N_{s} d_{s}^{3}}{6N_{c}LWT},
    \label{eq:mu_appr}
\end{gather}
where $\delta \eta$ provides the difference in packing fraction between the pure system of HBPs and the mixture of HBPs and HSs. Within the range of cuboid dimensions explored here,  $\delta \eta < 10^{-5}$ and it is therefore reasonable to assume that $\eta \sim \eta^{pure}$.

To equilibrate I and $\rm N_U$ phases, we performed standard Monte Carlo (MC) simulations in the canonical ensemble at the above-mentioned packing fractions and considered the systems at equilibrium when order parameters achieved a steady state value within reasonable statistical fluctuations. Given that all particles interact via hard-core potentials, MC moves are always accepted if they do not overlap. Overlap tests between two cuboids are based on the separating axes algorithm \cite{gottschalk1996} adapted to investigate the phase behaviour of colloidal HBPs with square cross section \cite{escobedo2005}. Overlap tests between spheres and cuboids are based on our recent Oriented Cuboid Sphere Intersection (OCSI) algorithm \cite{tonti2021}.

Uniaxial order parameters have been obtained from the diagonalization of the following tensor:
\begin{equation}
    \mathbbmss{Q}^{kk} = \frac{1}{2N_c} \sum_{i=1}^{N_{c}} \biggl[ 3 \Bigl( \vers{e}_{k,i} \otimes \vers{e}_{k,i} \Bigr) - \mathbbmss{I} \biggr],
\end{equation}
where the unit vector $\vers{e}_{k,i}$ indicates the orientation of $k = T,L,W$ for each cuboid $i$ and $\mathbbmss{I}$ is the unit tensor. The largest eigenvalue of $\mathbbmss{Q}^{kk}$, is the uniaxial order parameter relative to size $k$, here referred to as $U_k$, while the corresponding eigenvector, $\vect{d}_k$, indicates its preferential orientation. In an I phase, $U_T \approx U_W \approx U_L \approx 0$ and the vectors $\vect{d}_k$ are meaningless. By contrast, in an $\rm N_U$ phase, one eigenvalue is significantly larger than 0 and the associated eigenvector $\vers{n}$ defines the direction of the nematic director. The tensor  $\mathbbmss{Q}^{kk}$ also allows for the computation of the biaxial order parameters, which are anyway all close to zero for the systems studied here. To characterise the positional order of HBPs and spheres, we computed the radial distribution function, $g(r)$, in the I phases and pair correlation functions parallel, $g_{\parallel}(r)$, and perpendicular, $g_{\perp}(r)$, to the nematic director in the $\rm N_U$ phases. All of them have been obtained for cuboid-cuboid, sphere-sphere and sphere-cuboid distributions. The interested reader is referred to Refs.\ \cite{effran2020,Cuetos2004} for additional details on the calculation of parallel and perpendicular pair correlation functions.

\ctable[
 label = tab01:diff_hbp,
 pos = ht!,
 caption = {Translational and rotational diffusion coefficient at infinite dilution of the HBPs studied in this work.},
 star,
 width = \textwidth,
]{KKKKKKKK}{
}{
                                      \FL
 $W^*$ & $D^{tra}_{T} {D_0}^{-1}$ & $D^{tra}_{W} {D_0}^{-1}$ & $D^{tra}_{L} {D_0}^{-1}$   &
         $D^{rot}_{T} \tau$       & $D^{rot}_{W} \tau$       & $D^{rot}_{L} \tau$         \ML
 1     & $2.2 \cdot 10^{-2}$      & $2.2 \cdot 10^{-2}$      & $3.1 \cdot 10^{-2}$   &
         $1.1 \cdot 10^{-3}$      & $1.1 \cdot 10^{-3}$      & $2.3 \cdot 10^{-2}$   \NN
 8     & $9.4 \cdot 10^{-3}$      & $1.4 \cdot 10^{-2}$      & $1.5 \cdot 10^{-2}$   &
         $3.5 \cdot 10^{-4}$      & $3.6 \cdot 10^{-4}$      & $6.3 \cdot 10^{-4}$   \LL
}

As far as the dynamics is concerned, we generated time trajectories by employing the Dynamic Monte Carlo (DMC) simulation method \cite{Patti2012, Cuetos2015, Corbett2018, Daza2020}. In DMC simulations, the Brownian motion of colloidal particles in a fluid is modelled through stochastic displacements and rotations, whose timescale is set by the Einstein relation \cite{Einstein1905}. In particular, a random particle is selected and a trial translation is attempted if the particle is a sphere, $\vect{r}_{s,new} = \vect{r}_{s,old} + \delta \vect{r}_s$, or a rototranslation if it is a cuboid, namely $\vect{r}_{c,new} = \vect{r}_{c,old} + \delta \vect{r}_c$ for translation and $\vers{e}_{k,new} = \mathbbmss{R}_{TLW} \vers{e}_{k,old}$ for rotations. The rotation matrix $\mathbbmss{R}_{TLW}$ is employed to rotate the cuboids around their axes of orientation $\vers{e}_T,\vers{e}_L,\vers{e}_W$. The elementary displacements $\delta r_{k,s}$ of spheres along the reference axes $k = \vers{x},\vers{y},\vers{z}$ are uniformly sampled in the interval $[-\delta r_{k,max,s};\delta r_{k,max,s}]$. Similarly, the elementary displacements of cuboids, $\delta r_{k,c}$, are sampled in the interval
$[-\delta r_{k,max,c};\delta r_{k,max,c}]$, while their elementary rotations, $\delta \theta_{k,c}$, in 
$[-\delta \theta_{k,max,c};\delta \theta_{k,max,c}]$, for $k = T,L,W$.
The maximum displacements and rotations are defined using the Einstein relation given below:
\begin{gather}
    \delta r_{k,max,s} = \sqrt{2 D_s \delta t_{MC,s}} \quad k = \vers{x},\vers{y},\vers{z}, \\
    \delta r_{k,max,c} = \sqrt{2 {D^{tra}_{k,c}} \delta t_{MC,c}} \quad k = T,L,W, \\
    \delta \theta_{k,max,c} = \sqrt{2 {D^{rot}_{k,c}} \delta t_{MC,c}} \quad k = T,L,W,
\end{gather}
where $\delta t_{MC}$ is the MC timescale of the simulation, $D_s$ is the sphere diffusion coefficient at infinite dilution, whereas $D^{tra}_{k,c}$ and $D^{rot}_{k,c}$ are, respectively, the translational and rotational cuboid diffusion coefficients also at infinite dilution. In a DMC simulation of a binary mixture, the MC timescales of the two components are different, but the Brownian timescale is unique. These timescales are therefore related as follows:
\begin{equation}
    \mathscr{A}_{c} {\delta}t_{MC,c} = \mathscr{A}_{s} {\delta}t_{MC,s},
    \label{eq:dmc_eqv}
\end{equation}
where $\mathscr{A}_{c}$ and $\mathscr{A}_{s}$ are the acceptance rates of the attempted moves of cuboids and spheres, respectively. In practice, the MC timescale of either HBPs or HSs is kept constant to a given input value, while the other converges according to Eq.\,\ref{eq:dmc_eqv} by updating the acceptance rates. The Brownian timescale of the DMC simulation is linked to the acceptance rate of the components of the system and their MC timescales, according to Eq.\,\ref{eq:dmc_tbd}:
\begin{equation}
    t_{BD} = \frac{\mathscr{A}_{c}}{3} \mathscr{C}_{MC} {\delta}t_{MC,c} = \frac{\mathscr{A}_{s}}{3} \mathscr{C}_{MC} {\delta}t_{MC,s},
    \label{eq:dmc_tbd}
\end{equation}
where $\mathscr{C}_{MC}$ is the number of MC cycles performed, with 1 MC cycle corresponding
to $N_s + N_c$ attempted moves. A detailed theoretical discussion of the DMC method for mono- and multi-component systems is reported in Refs.\,\cite{Patti2012, Cuetos2015}.

To apply Eqs.\,(4-6), one needs to estimate the diffusion coefficient of spheres and cuboids at infinite dilution. The former is obtained from the Stokes-Einstein equation:
\begin{equation}
    D_s = \frac{D_0}{3 \pi {d_{s}}^*} \sim 1.061 D_0,
\end{equation}
where $D_0 \equiv T^2 \tau^{-1}$ is a diffusion constant, with $\tau$ the time unit. The diffusion tensor of HBPs at infinite dilution was estimated numerically using HYDRO++, an open-source software that calculates the solution properties of macromolecules and colloidal particles by approximating their shape and volume with an array of spherical beads of arbitrary size \cite{Carrasco1999, Garcia2007}. The diffusion tensor of biaxial particles is in principle  non-diagonal, implying the existence of a rototranslational coupling that is taken into account by a generalised version of the DMC method \cite{chiappini2020}. Nevertheless, the diffusion tensor that we obtained showed that its off-diagonal terms were at least 4 orders of magnitude smaller than the remaining terms and thus safely negligible. The translational and rotational diffusion coefficients at infinite dilution of prolate and oblate HBPs are listed in Table \ref{tab01:diff_hbp}. As in typical Brownian Dynamics and Langevin Dynamics simulations, Hydrodynamic Interactions (HI) are neglected also in DMC simulations. The effect of HI in BD simulations of crowded suspensions has been evaluated for binary mixtures of Lennard-Jones spheres of different sizes and results show that HI interactions slow down the dynamics of the spheres without modifying the qualitative features of the diffusion \cite{Kwon2014}.

In all DMC simulations, we set the MC time step of spheres in the range $\delta t_{MC,s} = \{5 \times 10^{-5}; 10^{-2} \} \tau$. The smallest time step determined a maximum displacement approximately equal to $d_s^*/10$, whereas the largest time step has been set in order to reproduce the same conditions as those recently studied in pure systems of HBPs \cite{cuetos2020}. We first performed a preliminary simulation of $10^4$ MC cycles to find the time step of cuboids according to Eq.\,(7), that is $\delta t_{MC,c} = {\mathscr{A}_s}^{(10)} \delta t_{MC,s} / {\mathscr{A}_c}^{(10)}$, averaging every 10 MC cycles. A summary of the systems investigated with the corresponding time steps are listed in Table \ref{tab02:summ}.

\ctable[
 label = tab02:summ,
 pos = ht!,
 caption = {Systems investigated in this work and associated simulation parameters. $N_s$ and $N_c$ refer to the number of HSs and HBPs, respectively; $d_s^*$ is the diameter of spherical tracers; $L^*$ and $W^*$ are the reduced cuboid length and width, respectively; $\eta^{pure} \approx \eta$ is the system packing fraction; $\delta t_{MC,s}$ and $\delta t_{MC,c}$ are the MC time steps of HSs and HBPs, respectively; and $\mathscr{A}_s$ and $\mathscr{A}_c$ are the MC acceptance rates of HSs and HBPs, respectively. The symbols $\rm N_U^+$ and $\rm N^-_U$ refer to uniaxial prolate and oblate nematic phases, respectively. The systems are indexed according to the shape of the HBPs and the packing fraction. Systems that have the same settings, but different input MC time step, are given the same index.},
 star,
 width = \textwidth,
]{cccccccKKKKK}{
}{
                                      \FL
 Systems   & $N_{s}$ & $N_{c}$ & ${d_{s}}^*$ & $L^*$ & $W^*$ & $\eta^{pure}$ & Phase       &
         $\delta t_{MC,s}$     & $\delta t_{MC,c}$     & $\mathscr{A}_s$ & $\mathscr{A}_c$ \ML
 $S_1$     & 200     & 2000    & 0.1         & 1     & 12    & 0.15         & I           &
        $5.000 \cdot 10^{-5}$  & $5.015 \cdot 10^{-5}$ & 0.998           & 0.995 \NN
 $S_1$     & 200     & 2000    & 0.1         & 1     & 12    & 0.15         & I           &
         $1.000 \cdot 10^{-2}$ & $1.043 \cdot 10^{-2}$ & 0.972           & 0.932 \NN
 $S_2$     & 200     & 2000    & 0.1         & 1     & 12    & 0.34         & $\rm N_U^+$ &
         $5.000 \cdot 10^{-5}$ & $5.034 \cdot 10^{-5}$ & 0.994           & 0.987 \NN
 $S_2$     & 200     & 2000    & 0.1         & 1     & 12    & 0.34         & $\rm N_U^+$ &
         $1.000 \cdot 10^{-2}$ & $1.109 \cdot 10^{-2}$ & 0.912           & 0.822 \NN
 $S_3$     & 200     & 2000    & 0.1         & 8     & 12    & 0.15         & I           &
         $5.000 \cdot 10^{-5}$ & $5.000 \cdot 10^{-5}$ & 0.999           & 0.999 \NN
 $S_3$     & 200     & 2000    & 0.1         & 8     & 12    & 0.15         & I           &
         $1.000 \cdot 10^{-2}$ & $1.006 \cdot 10^{-2}$ & 0.984           & 0.979 \NN
 $S_4$     & 200     & 2000    & 0.1         & 8     & 12    & 0.34         & $\rm N_U^-$ &
         $5.000 \cdot 10^{-5}$ & $5.013 \cdot 10^{-5}$ & 0.997           & 0.995 \NN
 $S_4$     & 200     & 2000    & 0.1         & 8     & 12    & 0.34         & $\rm N_U^-$ &
         $1.000 \cdot 10^{-2}$ & $1.031 \cdot 10^{-2}$ & 0.952           & 0.924 \LL
}

We characterised the dynamics of the systems investigated by computing the mean square displacement (MSD), the non-Gaussian parameter (NGP), the apparent exponent of the generalized Einstein relation and the self-part of the Van-Hove distribution functions (s-VHF), averaging out over 90 different uncorrelated trajectories. All these properties were computed for both the HBPs and the HSs and for different components: the 3D position of the particles, i.e. $\vect{r}_{tot}$ with dimensionality $d=3$, the components in the direction parallel and perpendicular to the nematic director in the $\rm N_U$ phases, i.e. $\vect{r}_{\parallel}$ with dimensionality $d=1$ and $\vect{r}_{\perp}$ with $d=2$, respectively, and along the three box axes in the I phases, $\vect{r}_{x},\vect{r}_{y},\vect{r}_{z}$, with $d=1$. In the following, we use the symbol ${\lambda} = \{tot,\parallel,\perp,x,y,z\}$ to indicate that the dynamical properties of a given observable have been estimated in 1, 2 and 3 dimensions. The MSD is defined as the ensemble average of the particle displacement from their original position at time 0:
\begin{equation}
    \Big \langle \Delta r^2_{\lambda} (t) \Big \rangle  = \frac{1}{N} \Bigg \langle \sum_{i=1}^{N} \big\| \vect{r}_{\lambda,i}(t) - \vect{r}_{\lambda,i}(0) \big\|^2 \Bigg \rangle.
\end{equation}
The diffusion coefficients at long timescales are obtained from the MSD as follows:
\begin{equation}
\label{eq11}
    D_{\lambda,long} = \frac{1}{2 d t}\lim_{t\to\infty} \langle \Delta r^2_{\lambda} (t) \rangle.
\end{equation}

Considering the generalised Einstein relation, where the MSD can have a nonlinear dependence on time, i.e. $\langle \Delta r^2_{\lambda} \rangle = 2dD_{\lambda} t^{\beta_\lambda}$, we can define the apparent exponent $\beta_\lambda$ as follows \cite{Ciesla2014}:
\begin{gather}
    \label{eq12}
    \ln{\langle \Delta r^2_{\lambda} \rangle} = \ln{(2dD_{\lambda})} + \beta_\lambda \ln{t}, \\
    \beta_\lambda = \frac{d\ln{\langle \Delta r^2_{\lambda} \rangle}}{d\ln{t}},
\end{gather}
where $\beta_\lambda$ describes deviations from a linear dependence of the MSD on time. 

The probability distribution of particle displacements at time $t$, given their position at time 0 can be defined as
\begin{equation}
    \nu_{d}G_{s,\lambda}(r,t) = \frac{1}{N} \Bigg \langle \sum_{i=1}^{N} \delta \bigg (r - \big \| \vect{r}_{\lambda,i}(t) - \vect{r}_{\lambda,i}(0) \big \| \bigg ) \Bigg \rangle,
\end{equation}
where $\nu_{d}G_{s,\lambda}(r,t)$ represents the probability that a particle $i$ has displaced a distance $r$, in 1, 2 or 3 dimensions, from its initial position at time $t$. In particular, $G_{s,\lambda}$ are the s-VHFs, and $\nu_{d}$ a normalization factor:
$\int_{0}^{\infty} G_{s,\lambda}(r_{\lambda},t) \nu_{d} dr_{\lambda} = 1$, where $\nu_1 = 1$ for the distributions parallel to the nematic director in the $\rm N_U$ phases or along $\vers{x},\vers{y},\vers{z}$ in the $\rm I$ phases, $\nu_2 = 2\pi r$, and $\nu_3 = 4\pi r^2$. If $G_{s,\lambda}$ are Gaussian-distributed, the s-VHFs can be approximated as:
\begin{equation}
    G_{s,\lambda}(r,t) = \frac{1}{\sqrt{\big( 4 \pi D_{\lambda,t} t \big)^d}} \exp{\Bigg( -\frac{r^2}{4D_{\lambda,t} t}\Bigg)},
    \label{eq:svh_appr}
\end{equation}
where $D_{\lambda,t} = \langle \Delta r^2_{\lambda} (t) \rangle / 2dt$ is the instantaneous diffusion coefficient along the direction $\lambda$. It has been recently showed that, if the diffusion of the particle is anisotropic, Eq.\,(\ref{eq:svh_appr}) does not correctly estimate (deviations from) the Gaussianity of particle displacements in all the 3D space, i.e. when $\lambda=tot$, and a Gaussian distribution with an ellipsoidal symmetry has been proposed \cite{cuetos2018}.

Finally, the non-Gaussian parameter (NGP) expresses deviations from the expected Gaussian distribution of the displacements. A general formula for the NGP is:
\begin{equation}
    \alpha_{2,d} = \frac{\langle \Delta r^4_{\lambda} (t) \rangle}{c_{2,d}\langle \Delta r^2_{\lambda} (t) \rangle^2} - 1.
    \label{eq:alpha_corr}
\end{equation}
The constant $c_{2,d}$ depends on the equivalence between the fourth moment, $\langle \Delta r^4_{\lambda} (t) \rangle$, and second moment, $\langle \Delta r^2_{\lambda} (t) \rangle$, of the Gaussian distribution of the displacements. For a perfect isotropic system, where there is no preferential direction for diffusion, the Gaussian distribution is spherical and the constant $c_{2,d} = (1+ 2/d)$, with $d=1,2,3$. If the system shows anisotropy and particles diffuse preferentially along a certain direction, the parameter $c'_{2,3}$ for a displacement in 3D space reads \cite{cuetos2018}
\begin{equation}
    c'_{2,3} = \frac{3D^2_{\parallel,t}+8D^2_{\perp,t}+4D_{\parallel,t}D_{\perp,t}}{D^2_{\parallel,t}+4D^2_{\perp,t}+4D_{\parallel,t}D_{\perp,t}}.
    \label{eq17}
\end{equation}
%


\section{Results and Discussion}

In this section, we study the dynamical properties of globular macromolecules immersed in colloidal suspensions of cuboids. The former are modelled as hard spherical tracers, while the latter as hard boards. Following standard MC simulations in the canonical ensemble, the complete set of uniaxial order parameters of the equilibrated systems were calculated and their values are summarised in Table~\ref{tab03:op}.

\ctable[
 label = tab03:op,
 pos = ht!,
 caption = {Average uniaxial order parameters relative to $T,L,W$ of oblate and prolate HBPs in $\rm I$ ($\eta=0.15$) and $\rm N_U$ ($\eta=0.34$) phases. The results reported are obtained from standard MC simulations in the canonical ensemble, after equilibration of the systems and before running DMC simulations. Absolute errors are lower than $5 \times 10^{-3}$.},
 width = \columnwidth,
]{KKKKK}{
}{
                                      \FL
 Systems   & Phase       & $U_T$   & $U_W$   & $U_L$ \ML
 $S_1$     & I           & 0.020   & 0.020   & 0.025 \NN
 $S_2$     & $\rm N_U^+$ & 0.255   & 0.256   & 0.962 \NN
 $S_3$     & I           & 0.031   & 0.021   & 0.023 \NN
 $S_4$     & $\rm N_U^-$ & 0.943   & 0.250   & 0.252 \LL
}

\begin{figure}[ht!]
    \centering
    \includegraphics[width=\columnwidth]{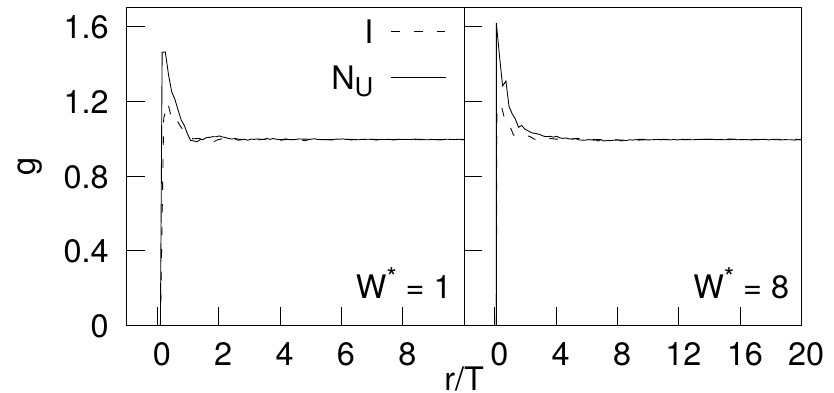}
    \caption{HS - HS Radial distribution functions in $\rm I$ (dashed lines) and $\rm N_U$ (straight lines) phases of prolate (left panel) and oblate (right panel) HBPs. The flat profile of the radial distribution functions from short radius prove the uncorrelation of the relative position of the spherical tracers for all the systems investigated.}
    \label{fig02:GOFR_SS}
\end{figure}

\begin{figure}[ht!]
    \centering
    \includegraphics[width=\columnwidth]{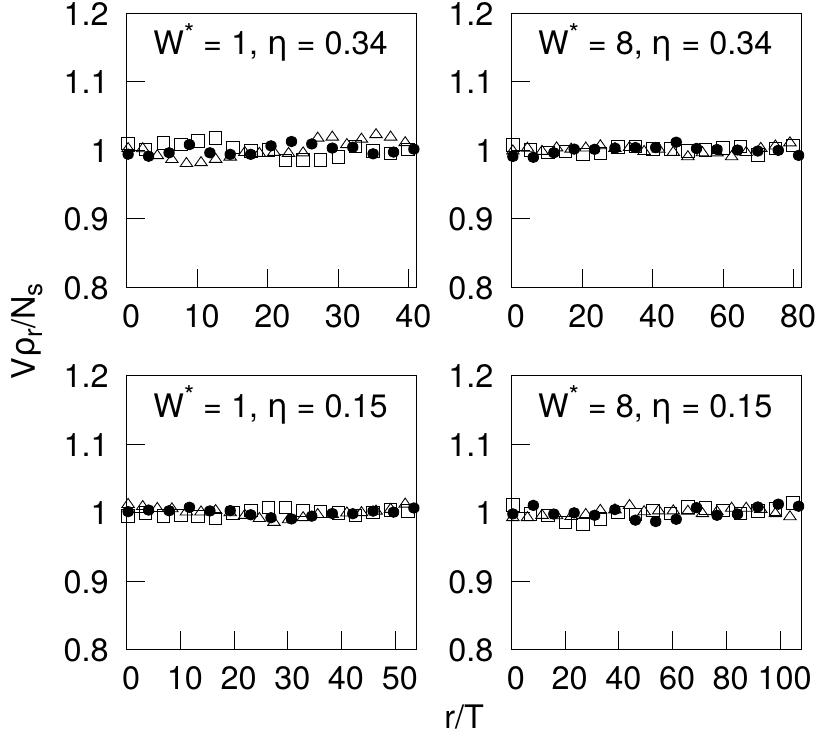}
    \caption{Profile of the average numeral density of HS in slabs, i.e. $\rho_r$, normalized by the numeral density of the HS in the entire box, i.e. $N_s / V$, for all the systems investigated. The slabs have volume $V^{2/3}\Delta r$, where $\Delta r$ is oriented along $\vers{x}$ ($\triangle$), along $\vers{y}$ ($\bullet$) and along $\vers{z}$ ($\Box$). Flat profiles prove that the spherical tracers are homogeneously distributed.}
    \label{fig01:SLICE}
\end{figure}

\noindent We also computed the biaxial order parameters and no significant long-ranged biaxiality was detected, in agreement with previous works on monodisperse HBPs \cite{Cuetos2017}. Additionally, inspection of the pair correlation functions and the snapshots of the equilibrated systems, respectively reported in Figs.\,S1, S2 and S3 of the Supporting Information, does not reveal long-ranged positional ordering of HBPs, thus suggesting that smectic phases are not formed. We also note that the presence of spherical tracers does not affect the morphology of the phases observed in pure systems of HBPs at the same packing fractions. In order to ensure that the positional distribution of the spherical tracers is homogeneous in the entire volume of the systems simulated, we computed the HS - HS radial distribution functions (Fig.\,\ref{fig02:GOFR_SS}) and the density profile of the HS along the $\vers{x},\vers{y},\vers{z}$ directions of the reference axes (Fig.\,\ref{fig01:SLICE}). The density profile was computed as the average number of HS found in a slab, i.e. $\rho_r = \langle N_s \rangle_{\rm slab} / V_{\rm slab}$, with $V_{\rm slab} = V^{2/3} \Delta r$ for $\Delta r$ oriented in the three directions of the box frame, normalised by the density of the tracers all over the box, i.e. $N_s / V$. All the one-particle distributions along the reference axes are flat and equal to the numeral density of the tracers, clearly proving that the tracers are homogeneously distributed. In addition, all the HS - HS radial distribution functions decay to unity at short distances, proving that the position of the tracers is completely uncorrelated in all the systems.

\subsection{Dynamical properties}

\begin{figure}[ht!]
    \centering
    \includegraphics[width=\columnwidth]{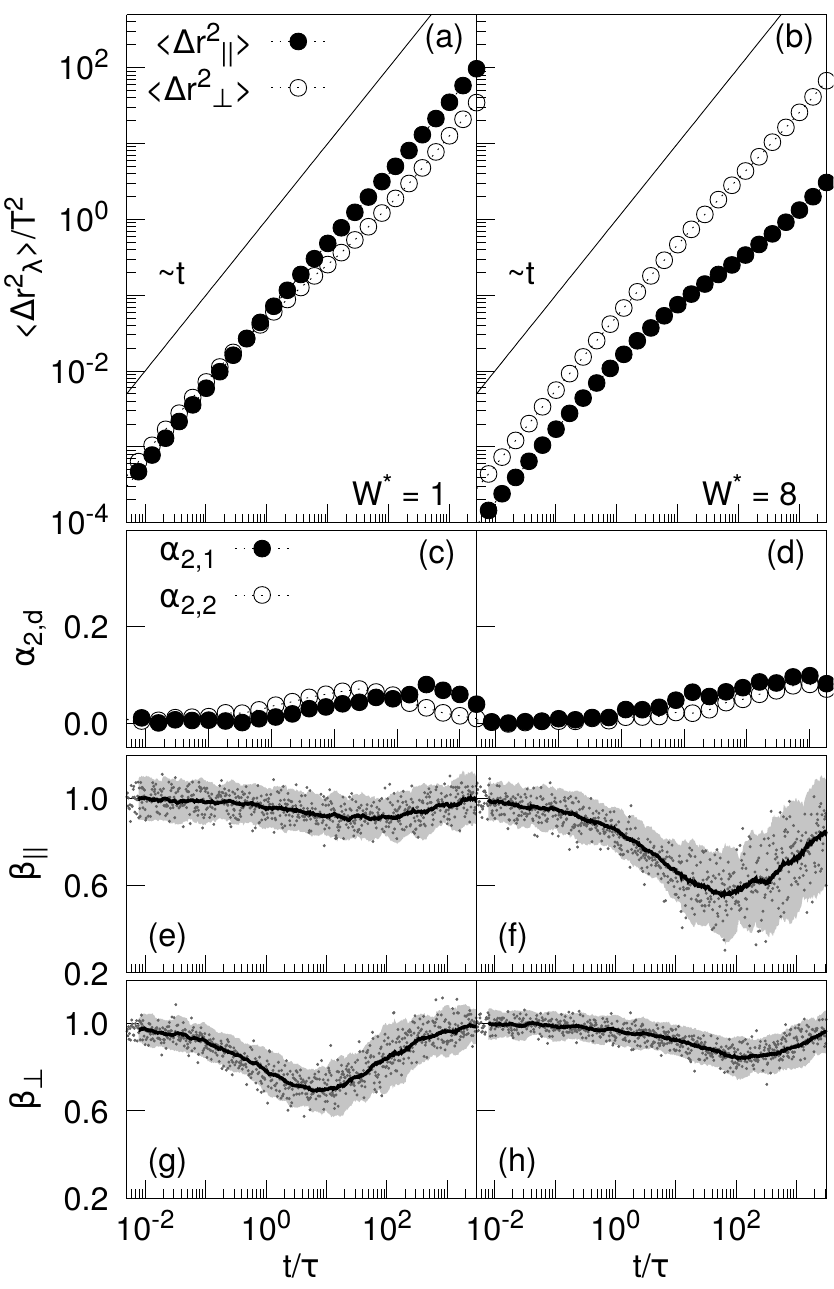}
    \caption{MSDs (a,b) and NGPs (c,d) of HBPs in the direction parallel ($\bullet$) and perpendicular ($\circ$) to $\vers{n}$ in prolate (left) and oblate (right) nematic phases, at packing fraction $\eta = 0.340$. Frames (e) to (h) report the instantaneous values (dots), average (thick line) and twice its standard deviation (gray-shaded area) of the apparent exponents $\beta_\parallel$ and $\beta_\perp$. Solid lines in panels (a) and (b) indicate the expected dependence of MSD on time in the Fickian regime.}
    \label{fig03:MSDN_c}
\end{figure}

Fig.\,\ref{fig03:MSDN_c} depicts the MSD, NGP and apparent exponent of  prolate ($W^*=1$) and oblate ($W^*=8$) HBPs in $\rm{N_U}$ phases in the directions parallel and perpendicular to the nematic director. The dynamics of HBPs is anisotropic as the distinct dependence of $\langle \Delta {r}^2_\parallel \rangle$ and $\langle \Delta {r}^2_\perp\rangle$ on time reveals. However, an analogous sequence of dynamical regimes is observed for all the cases studied. At short times, HBPs diffuse within the cage formed by nearby particles, rattling around their initial location and not yet interacting with their neighbours. At this stage, the MSD is linear with time. At slightly larger times, the effect of this cage fully develops as diffusion is slowed down due to the collisions with the surrounding particles, resulting in a nonlinear MSD with time. Finally, the expected Brownian motion sparks at sufficiently long time scales, when the MSD recovers its linear dependence on time. 

\noindent According to Fig.~\ref{fig03:MSDN_c}, the MSD perpendicular to $\vers{n}$ is larger than its parallel counterpart at short times. This is caused by the higher dimensionality in the displacements perpendicular to $\vers n$ ($d=2$) compared to those parallel to it ($d=1$). At intermediate times, a sub-diffusive behaviour is observed for both prolate and oblate HBPs, which depends on the preferential orientation of the nematic phases and particle anisotropy. While HBPs in the $\rm N^+_U$ phase ($W^{*} = 1$) are predominantly oriented along the axis parallel to their length $L$ and move preferentially along $\vers{n}$ (solid circles in Fig.~\ref{fig03:MSDN_c}(a)), particles in the $\rm N^-_U$ phase ($W^{*} = 8$) are mostly oriented along their thickness $T$ and their motion along $\vers{n}$ is hampered (solid circles in Fig.~\ref{fig03:MSDN_c}(b)). Prolate HBPs also experience a similar slow down in planes perpendicular to $\vers{n}$, but this effect is significantly less pronounced. Finally, at long times, the particles enter a new diffusive regime in which $\langle\Delta r^2_\lambda\rangle$ recovers its linearity with time. All these observations can be further appreciated by the evolution of the apparent exponents $\beta_\parallel$ and $\beta_\perp$ in Figs.\,\ref{fig03:MSDN_c}(e) and \ref{fig03:MSDN_c}(g) for prolate HBPs and Figs.\,\ref{fig03:MSDN_c}(f) and \ref{fig03:MSDN_c}(h) for oblate HBPs, respectively. In the direction along which the particle motion is especially hampered,  $\beta_\lambda$ shows larger deviations from 1 (the Fickian-like value), corresponding to a sub-diffusive behaviour and the formation of temporary cages. The most relevant sub-diffusive regime is observed in systems of oblate HBPs in the direction parallel to $\vers{n}$, where $\beta_\parallel \sim 0.6$ at $t/\tau \sim 100$. The NGP of prolate and oblate HBPs are shown in panels (c) and (b) of Fig.~\ref{fig03:MSDN_c}, respectively, along $\vers{n}$ and perpendicularly to it. Although a subtle growth of the NGP is observed in all systems, its maximum value is relatively low as it does not exceed 0.1. For comparison, the NGP of hard spherocylinders in smectic phases was reported to be between 3 at $\eta=0.508$ and 7 at $\eta=0.557$ in the direction of $\vers{n}$ \cite{Patti2009}. This suggests that the dynamics in nematic LC phases of HBPs is basically  Gaussian-like as also reported in a recent work \cite{cuetos2020}. To better understand the effect of particle geometry on the mobility of HBPs in nematic LCs, we have calculated the long-time translational diffusion coefficients, which are listed in Table~\ref{tab04:diff_long_C}. One can observe that, as the particle width increases, the diffusion coefficient parallel to $\vers n$ decreases significantly. Upon increasing $W^*$, the probability of HBPs to collide with their neighbours increases as well, which in turn slows down the overall mobility. The opposite effect occurs for diffusion coefficients perpendicular to $\vers n$ where a small increment is observed when the width of the HBPs increases. Our MSDs and diffusion coefficients are in excellent quantitative agreement with those reported by Cuetos and Patti in systems of pure HBPs \cite{cuetos2020}. This agreement suggests that the presence of spherical tracers does not affect the dynamics of HBPs in $\rm N_U^+$ or $\rm N_U^-$ phases. 

\ctable[
 label = tab04:diff_long_C,
 pos = ht!,
 caption = {Long-time translational diffusion coefficients of HBPs parallel and perpendicular to $\vers{n}$ in $\rm N_U$ phases of HBPs, at packing fraction $\eta=0.34$ for both systems. The absolute errors are lower than half of the last significant digit.},
 width = \columnwidth,
]{KKK}{
}{
                                      \FL
 Systems & $D_{\parallel,long}{D_0}^{-1}$ & $D_{\perp,long}{D_0}^{-1}$ \ML
 $S_2$   & $1.7 \cdot 10^{-2}$            & $3.1 \cdot 10^{-3}$        \NN
 $S_4$   & $4.8 \cdot 10^{-4}$            & $5.9 \cdot 10^{-3}$        \LL
}

\begin{figure}[ht!]
    \centering
    \includegraphics[width=\columnwidth]{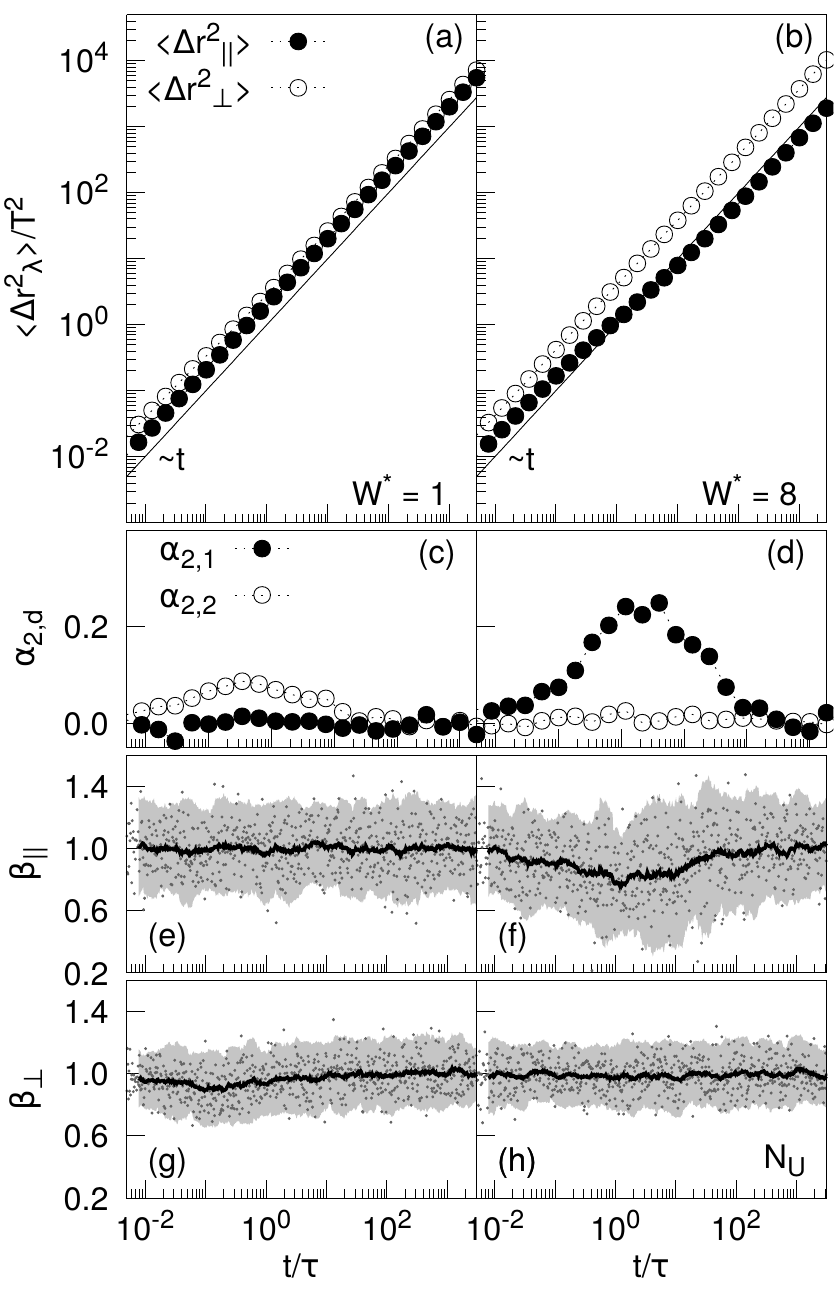}
    \caption{MSDs (a,b) and NGPs (c,d) of HSs in the direction parallel ($\bullet$) and perpendicular ($\circ$) to $\vers{n}$ in prolate (left) and oblate (right) nematic phases of HBPs, at packing fraction $\eta = 0.340$. Frames (e) to (h) report the instantaneous values (dots), average (thick line) and twice its standard deviation (gray-shaded area) of the apparent exponents $\beta_\parallel$ and $\beta_\perp$. Solid lines in panels (a) and (b) indicate the expected dependence of MSD on time in the Fickian regime.}
    \label{fig04:MSDN_s}
\end{figure}

In the light of these considerations on the dynamics of HBPs, we now examine the dynamics of the dispersed HSs. In Fig.~\ref{fig04:MSDN_s}, we show their MSDs, NGPs and apparent exponents in $\rm N_U$ phases of prolate (left panels) and oblate (right panels) HBPs. Due to their relatively small size, spherical tracers are expected to explore the available space more effectively than HBPs. This is indeed confirmed by their MSDs, which are larger than those of HBPs at all time scales. These results are also in agreement with Brownian dynamics simulations of mixtures of spheres of different sizes \cite{Kwon2014, Hwang2016}. The MSDs of the spherical tracers have a behaviour similar to that observed for HBPs: a linear behaviour at short and long times, and a very soft, almost negligible sub-diffusive behaviour at intermediate times. As the geometry of HBPs changes from prolate (Fig.~\ref{fig04:MSDN_s}(a)) to oblate (Fig.~\ref{fig04:MSDN_s}(b)), the sub-diffusive regime of $\langle\Delta {r}^2_{\parallel}\rangle$ becomes more evident, while no changes are practically detected in $\langle\Delta {r}^2_{\perp}\rangle$. These tendencies can be better appreciated by analysing the NGPs in Figs.~\ref{fig04:MSDN_s}(c) and \ref{fig04:MSDN_s}(d). While the peak of $\alpha_{2,2}$ vanishes from $\rm N_U^+$ to $\rm N_U^-$, the peak of $\alpha_{2,1}$ increases, indicating more pronounced deviations from Gaussianity in nematic LCs of oblate HBPs. Interestingly, while the peak of $\alpha_{2,2}$ in the $\rm N_U^+$ phase occurs at $t/\tau\sim 0.4$ and is lower than 0.1, confirming the substantially Gaussian dynamics of tracers in prolate nematics, the peak of $\alpha_{2,1}$ in the $\rm N_U^-$ phase, observed at $t/\tau\sim 3.3$, is significantly larger and approximately equal to 0.3.  This result indicates that long-range ordering can have a relevant influence on macromolecular diffusion as not only does it establish preferential paths of mobility, but it can also affect the nature of these paths by inducing deviations from Gaussianity, at least at intermediate times. At long time scales, the NGPs decay to zero and, correspondingly, the MSDs end up displaying Fickian behaviour. The cage effect exerted by the neighboring HBPs determine the transient sub-diffusive regimes. This cage provides a barrier against diffusion of both HBPs and guest molecules. Due to their smaller size and increased mobility as compared to HBPs, the spherical tracers perceive the effect of the surrounding cage at shorter times and over a shorter time interval than HBPs, as the position and broadness of the peak of the NGPs in panels (c) and (d) of Fig.\,\ref{fig03:MSDN_c} with respect to those in Fig.\,\ref{fig04:MSDN_s} reveal. The analysis of the HSs' apparent exponents in Fig. \ref{fig04:MSDN_s}(e-h) suggests that the tracers maintain a Fickian-like diffusion at all times. The only exception is detected in the $\rm N^-_U$ phase for diffusion along $\vers n$, where $\beta_{\parallel}<1$ for $t<10^2\tau$. Such a temporary non-Fickian and non-Gaussian dynamics is also observed in the host phase (see Fig.\,\ref{fig03:MSDN_c}(f)). The main difference between the dynamics of host and guest particles, in this case, is in the peak of $\alpha_{2,1}$, which is larger for HSs than for HBPs. This is not surprising if one considers that HSs are mostly surrounded by oblate HBPs whose dynamics, relatively slow in the direction of $\vers n$, slows down the diffusion of HSs, rather abruptly, after the initial diffusive regime. Parallel and perpendicular long-time diffusion coefficients of HSs, obtained with Eq.\,\ref{eq11}, are listed in Table \ref{tab05:diff_long_s}.

\ctable[
 label = tab05:diff_long_s,
 pos = ht!,
 caption = {Long-time translational diffusion coefficients of HSs parallel and perpendicular to $\vers{n}$ in $\rm N_U$ phases of HBPs, at $\eta=0.34$. Absolute errors are smaller than half of the last significant digit.},
 width = \columnwidth,
]{KKK}{
}{
                                      \FL
 Systems & $D_{\parallel,long}{D_0}^{-1}$ & $D_{\perp,long}{D_0}^{-1}$ \ML
 $S_2$   & $9.9 \cdot 10^{-1}$            & $6.4 \cdot 10^{-1}$        \NN
 $S_4$   & $3.5 \cdot 10^{-1}$            & $9.0 \cdot 10^{-1}$        \LL
}

\begin{figure}[ht!]
    \centering
    \includegraphics[width=\columnwidth]{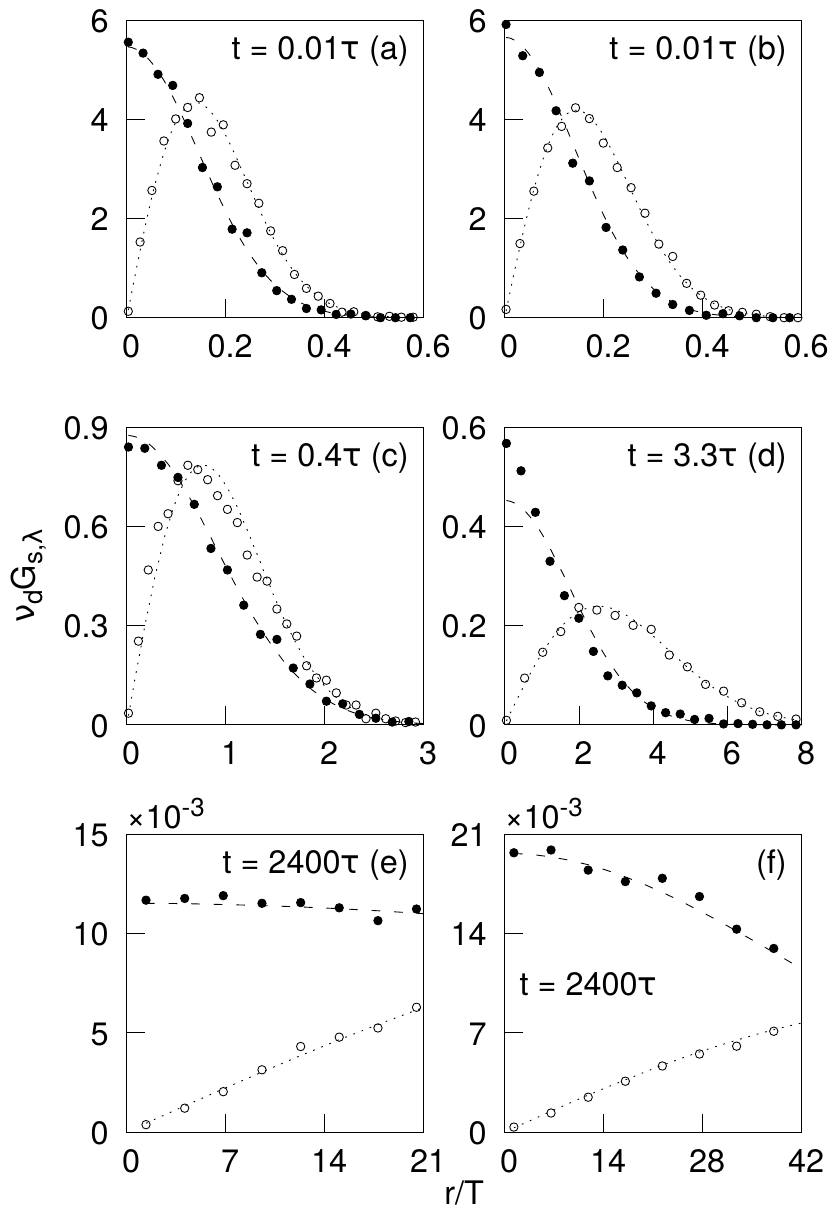}
    \caption{Self-Van Hove distributions of HSs in nematic phases of HBPs with $W^*=1$ (panels (a), (c) and (e)) and $W^*=8$ (panels (b), (d) and (f)), at $\eta=0.340$, at different times. Histograms obtained from simulations are plotted as points for the components parallel ($\bullet$) and perpendicular ($\circ$) to $\vers{n}$. Dashed and dotted lines refer to the theoretical Gaussian distributions obtained from Eq.~\ref{eq:svh_appr} for the directions parallel and perpendicular to $\vers{n}$, respectively.}
    \label{fig05:SVHN_S}
\end{figure}

While, in prolate nematics, spherical tracers diffuse faster in the direction of $\vers n$ than perpendicularly to it, with $D_{\parallel,long} \approx 1.55 D_{\perp,long}$, in oblate nematics this tendency changes dramatically as the parallel diffusion becomes significantly slower than the perpendicular diffusion, with $D_{\parallel,long} \approx 0.39 D_{\perp,long}$. Such a dependence of the dynamics of HSs on the geometry of the host HBPs and, ultimately, on the symmetry of the nematic phase, reveals a close analogy with the diffusion of HBPs, which is faster along $\vers n$ in the $\rm N_U^+$ phase and perpendicularly to it in the $\rm N_U^-$ phase (see Table~\ref{tab04:diff_long_C}). This result suggests that the anisotropic mobility of guest HSs is closely controlled by the dynamics of host HBPs. We notice that the decoupling of the dynamics of our spherical tracers as a result of the space anisotropy determined by the host cuboids, exhibits common characteristics with the dynamics of apoferritin, a globular protein generally found in the intestinal membrane, in I and N suspensions of \textit{fd} viruses at different concentrations \cite{Kang2006, Kang2007}. The time evolution of the self-Van Hove functions of HSs in prolate and oblate nematic phases is presented, respectively, in the left and right frames of Fig.~\ref{fig05:SVHN_S}. For each phase, we show the s-VHFs at three different times: (\textit{i}) $t/\tau=0.01$ (frames (a) and (b)), being a representative time for short-time diffusion in both phases; (\textit{ii}) $t/\tau=0.4$ (frame (c)) and $t/\tau=3.3$ (frame (d)), which pinpoint the crossover from short- to long-time diffusion in prolate and oblate nematics, respectively; and (\textit{iii}) $t/\tau=2400$ (frames (e) and (f)), which is a representative time of the long-time diffusive regime. Parallel and perpendicular s-VHFs were also estimated with the Gaussian approximation of Eq.~\ref{eq:svh_appr} with $d=1$ and 2, respectively. These Gaussian distributions are also plotted in Fig.~\ref{fig05:SVHN_S} as dotted and dashed lines. As a general tendency, the distribution of displacements as obtained from simulations is practically Gaussian at short and long times, while small deviations are detected at intermediate times, especially in the $\rm N_U^-$ phase (frame (d) in Fig.~\ref{fig05:SVHN_S}). In this case, the Gaussian approximation significantly sub-estimates the probability of HSs to remain in their original position or very close to it at $t/\tau=3.3$. This implies that there are more \textit{slow} tracers than those a Gaussian distribution would predict. By contrast, the probability of observing \textit{fast} tracers, given by the tail of the s-VHFs at large distances, is approximated  very well by a Gaussian distribution across the six time decades simulated here and in both $\rm N_U^+$ and $\rm N_U^-$ phases. We also notice that, at long times, the Fickian-like dynamics of our model globular macromolecules is also Gaussian, again discarding the ubiquity of Fickian yet non-Gaussian diffusion in soft-matter systems \cite{cuetos2018, cuetos2020}. 

\begin{figure}[ht!]
    \centering
    \includegraphics[width=\columnwidth]{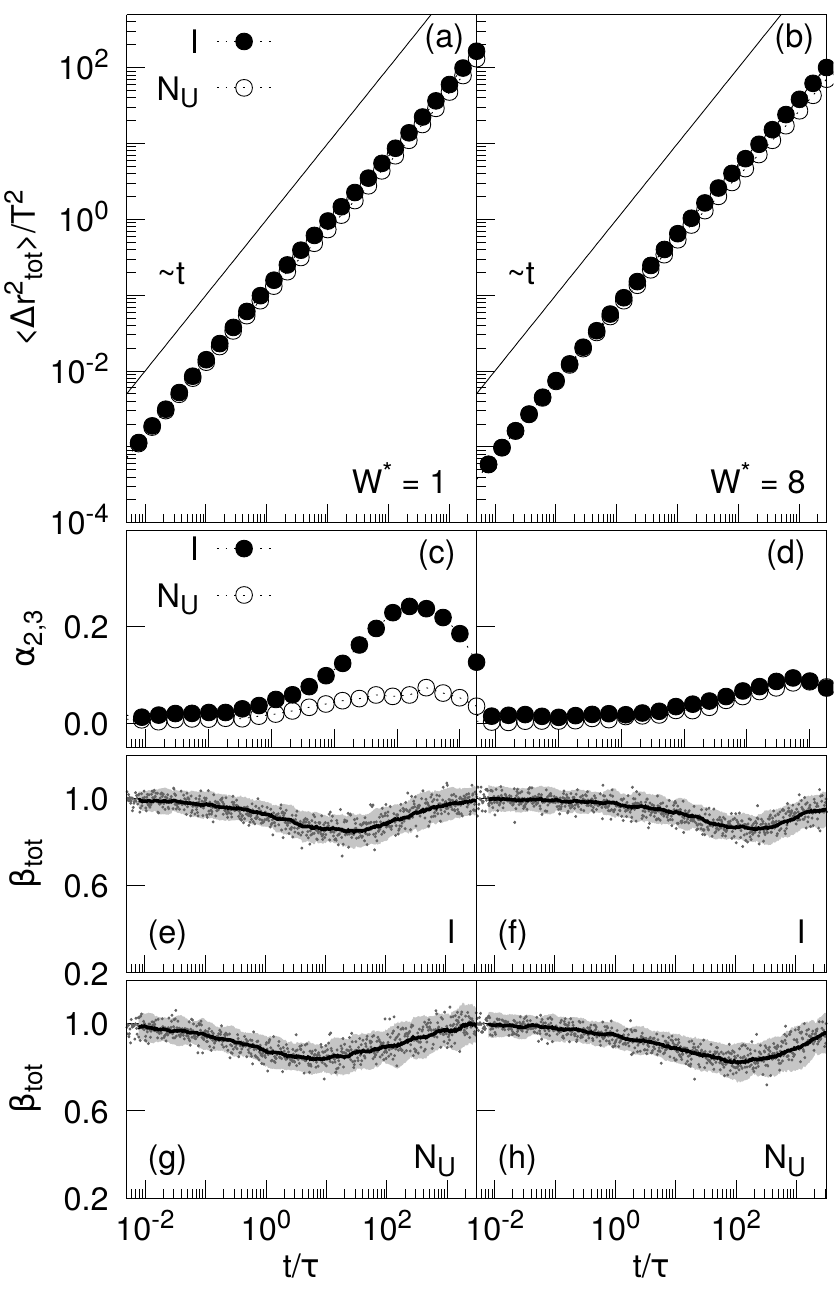}
    \caption{Total MSDs (a-b), NGPs (c-d) and apparent exponents (e-h) of prolate (left) and oblate (right) HBPs in $\rm I$ ($\bullet$), at $\eta = 0.150$, and $\rm N_U$ phases ($\circ$), at $\eta = 0.340$. Solid lines in panels (a-b) indicate the expected dependence of MSD on time in the Fickian regime. Frames (e-h) report the instantaneous values (dots), average (thick line) and twice its standard deviation (gray-shaded area) of the apparent exponent.}
    \label{fig06:MSDI_c}
\end{figure}

\begin{figure}[ht!]
    \centering
    \includegraphics[width=\columnwidth]{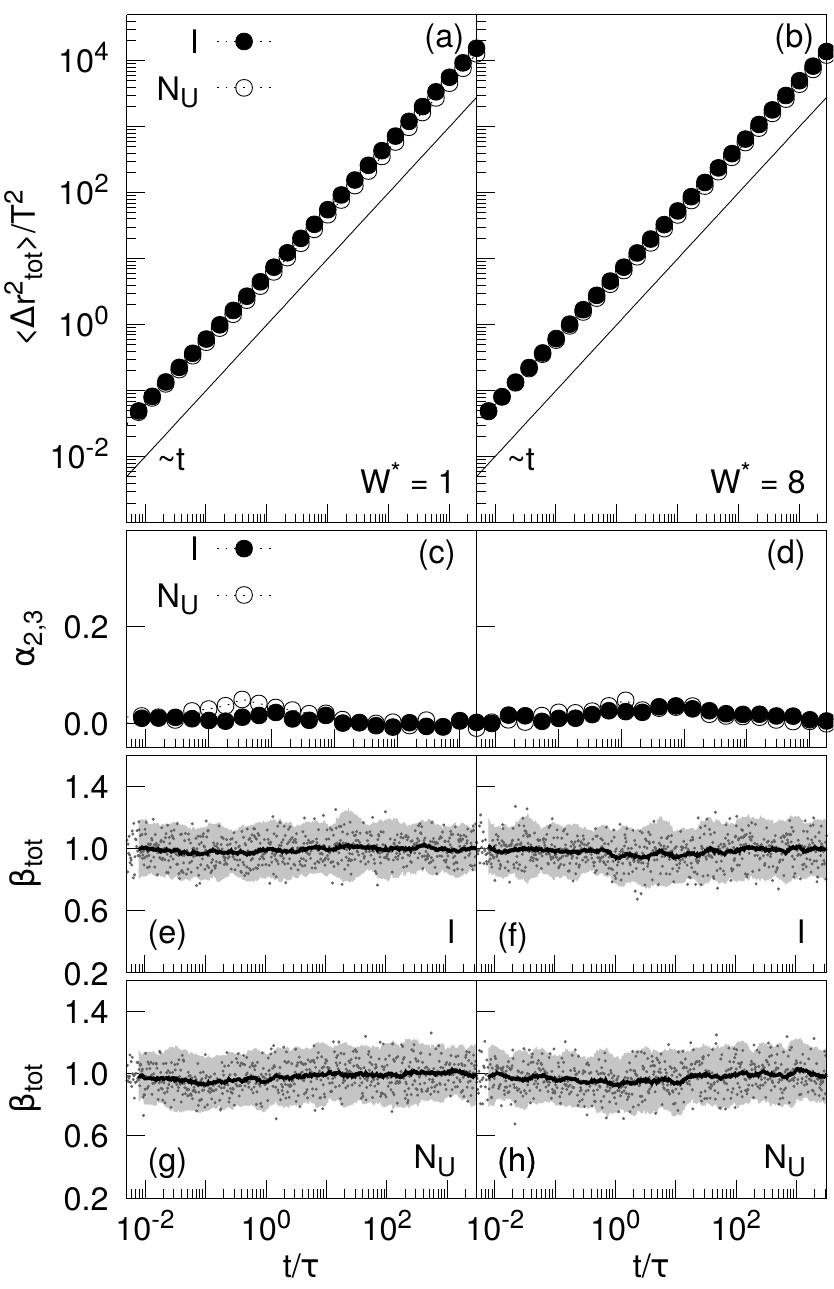}
    \caption{Total MSDs (a-b), NGPs (c-d) and apparent exponents (e-h) of HSs in $\rm I$ ($\bullet$) and $\rm N_U$ phases ($\circ$) of prolate (left frames) and oblate (right frames) HBPs. Both $\rm I$ phases have a packing fraction of $\eta = 0.150$, while both $\rm N_U$ phases have a packing fraction of $\eta = 0.340$. Solid lines in panels (a-b) indicate the expected dependence of MSD on time in the Fickian regime. Frames (e-h) report the instantaneous values (dots), average (thick line) and twice its standard deviation (gray-shaded area) of the apparent exponent.}
    \label{fig07:MSDI_s}
\end{figure}

To better understand how the long-range orientational ordering of nematics influences macromolecular diffusion, we have also investigated the dynamics of our tracers in $\rm I$ phases of HBPs. We firstly notice that the MSDs and NGPs of both host and guest species along the three space directions (shown in Figs.\,S4 and S5 of the Supplementary Material) reveal the same quantitative behaviour and confirm the absence of any preferential direction of motion. Consequently, it makes sense to limit our analysis only to the total MSD ($\rm \langle \Delta r^2_{tot} \rangle$) and NGP ($\alpha_{2,3}$). Both properties are reported, along with the total apparent exponent ($\rm \beta_{tot}$), in Figs.\,\ref{fig06:MSDI_c} and \ref{fig07:MSDI_s} for HBPs and HSs, respectively. More specifically, Fig.\,\ref{fig06:MSDI_c} displays these three dynamical properties in $\rm I$ and $\rm N_U$ phases of prolate (left frames) and oblate (right frames) HBPs. At short times, the dynamics of HBPs results to be unaffected by the degree of local ordering, if any, as the total MSDs in the $\rm I$ and $\rm N_U$ phases, shown in Figs.\,\ref{fig06:MSDI_c}(a) and \ref{fig06:MSDI_c}(b), are practically identical. This is not surprising as, at these short time scales, HBPs are still rattling around their original position and do not yet perceive the presence and degree of ordering of their neighbours. At longer time scales, the MSD slightly deviates from its linear behaviour with time and enters a sub-diffusive regime. At this stage, HBPs start to collide with each other and their dynamics slow down. Such a temporary cage effect is more effectively appreciated by analysing $\alpha_{2,3}$ (Fig.\,\ref{fig06:MSDI_c}(c-d)) and $\rm \beta_{tot}$ (Fig.\,\ref{fig06:MSDI_c}(e-h)). The apparent exponent presents very similar trends at $W^*=1$ and 8, with deviations from unity that do not depend on the background ordering. Because $\rm \beta_{tot}$ is a measure of the instantaneous Fickianity of the diffusion, these features and the analogies between the dynamics in $\rm I$ and $\rm N_U$ phases, are also observed in the MSDs at intermediate and long time scales. Especially intriguing is the analysis of the NGPs, reported in Fig.\,\ref{fig06:MSDI_c}(c) and \ref{fig06:MSDI_c}(d). Oblate HBPs show almost indistinguishable NGPs in both $\rm I$ and $\rm N_U^-$ phases and across the three time regimes, with a very small peak at $t/\tau \approx 500$. A very similar trend is also observed for prolate HBPs, but only in the $\rm N_U^+$ phase (empty circles in frame (c)). The NGP of prolate HBPs in the $\rm I$ phase exhibits a peak that is located at slightly earlier times ($t/\tau \approx 200$) and is surprisingly larger than any other. Because  deviations from Gaussianity usually increase with system packing \cite{Vorselaars2007, Belli2010, Matena2010, Patti2011}, it is quite remarkable to find a family of particles that inverts this tendency when compressed from the $\rm I$ to the $\rm N_U^+$ phase. We have analysed the origin of this atypical behaviour and found that it is most probably determined by the occurrence of randomly-oriented nematic-like clusters in the $\rm I$ phase. A detailed discussion is provided in Section 3.2. 

In Fig.~\ref{fig07:MSDI_s}, we report the total MSDs, NGPs and apparent exponents of the spherical tracers in $\rm I$ and $\rm N_U$ phases of prolate (left frames) and oblate (right frames) HBPs. The three sets of observables suggest a Fickian and Gaussian dynamics across the whole time scales and in both phases. These results might appear in contradiction with those reported in Fig.~\ref{fig04:MSDN_s}, showing sub-diffusive, non-Gaussian dynamics at intermediate time scales in the direction parallel to $\vers n$ in the $\rm N^-_U$ phase. Nevertheless, spherical tracers move preferentially in planes perpendicular to $\vers n$ as the difference between the MSDs measured in Fig.\,\ref{fig04:MSDN_s}(b) indicates. Consequently, although the dynamics along $\vers n$ deviates from Brownian diffusion over a relevant period of time, these deviations are averaged out when one computes $\rm \langle \Delta r^2_{tot} \rangle$, $\alpha_{2,3}$ and $\rm \beta_{tot}$. It is only by decoupling the HSs' dynamics according to the background anisotropy imposed by the host phase that we are able to detect the full picture and appreciate elements that otherwise would go underground. The total long-time diffusion coefficients, $D_{tot,long}$, of  HBPs and HSs in the $\rm I$ and $\rm N_U$ phases are calculated by using Eq.~\ref{eq11} with $d=3$ and summarised in Table \ref{tab06:diff_long_tot}. Increasing the system packing has a negative effect on the mobility of HSs and HBPs by substantially decreasing the free space available for diffusion. Consequently, the long-time diffusion coefficient of HBPs in $\rm N_U$ phases is $\sim 20-30\%$ smaller than that measured in the $\rm I$ phases (systems $S_1$ \textit{vs} $S_2$ and $S_3$ \textit{vs} $S_4$). A slightly less strong effect is observed in the case of the HSs, whose long-time diffusivities in the $\rm N_U$ phases are  $\sim 10-20\%$ smaller than those in the $\rm I$ phase.

\ctable[
 label = tab06:diff_long_tot,
 pos = ht!,
 caption = {Total long-time translational diffusion coefficients of HBPs and HSs, in $\rm I$ ($\eta=0.15$ for systems $S_1$ and $S_3$) and $\rm N_U$ ($\eta=0.34$ for systems $S_2$ and $S_4$) phases of HBPs. Absolute errors are smaller than half of the last significant digit.},
 width = \columnwidth,
]{KKK}{
}{
                                      \FL
 \multirow{2}{*}{Systems} & $D_{tot,long}{D_0}^{-1}$ & $D_{tot,long}{D_0}^{-1}$ \NN
  & HBPs & HSs \ML
 $S_1$   & $9.7 \cdot 10^{-3}$            & $9.3 \cdot 10^{-1}$             \NN
 $S_2$   & $7.7 \cdot 10^{-3}$            & $7.6 \cdot 10^{-1}$            \NN
 $S_3$   & $5.7 \cdot 10^{-3}$            & $8.1 \cdot 10^{-1}$             \NN
 $S_4$   & $4.1 \cdot 10^{-3}$            & $7.2 \cdot 10^{-1}$             \LL
}

\subsection{Cluster formation in isotropic phases of HBPs}

\begin{figure}[!ht]
    \centering
    \includegraphics[width=\columnwidth]{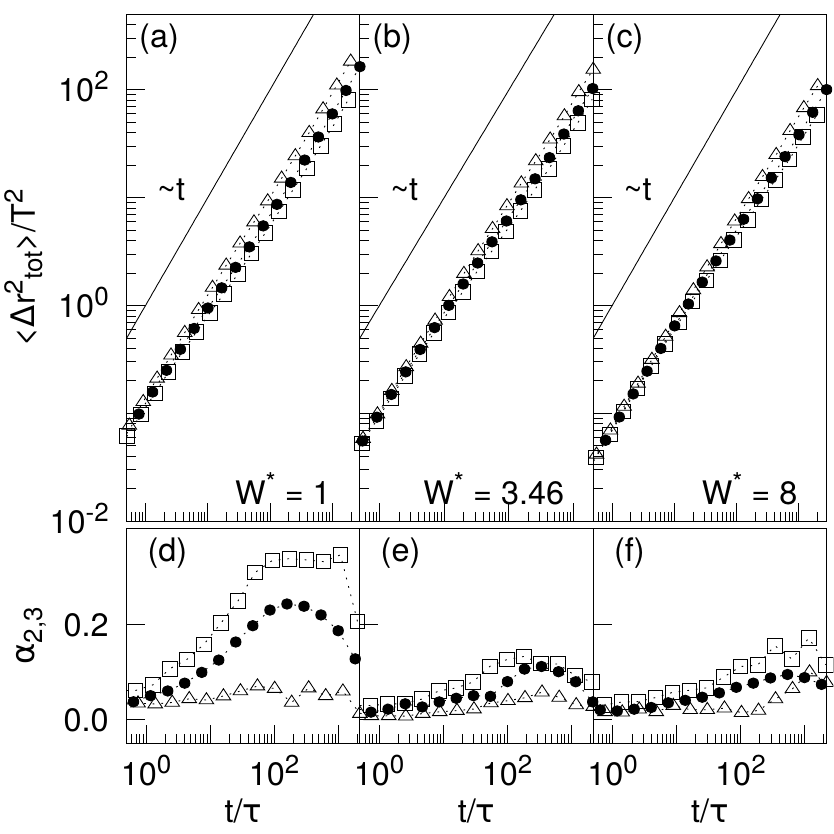}
    \caption{Total MSDs (top frames) and corresponding NGPs (bottom frames) in $\rm I$ phases of pure HBPs of different shape anisotropy. Empty triangles ($\triangle$), solid circles ($\bullet$) and empty squares ($\Box$) refer to packing fractions $\eta = 0.07$, 0.15 and 0.20, respectively. Solid lines in frames (a-b) indicate the expected dependence of MSD on time in Fickian diffusion.}
    \label{fig08:MSD_CL}
\end{figure}

To throw light on the origin of the intriguing non-Gaussian dynamics of prolate HBPs observed in the $\rm I$ phase at intermediate time scales, we first compare the results obtained in isotropic mixtures at $\eta = 0.15$ with those of pure systems of HBPs at $\eta = \{0.07,0.15,0.20\}$. At each of these packing fractions, we kept the total number of cuboids equal to $N = 2000$ and particle thickness and length constant to $T$ and $12T$, respectively, and varied particle width. In particular, we confirmed $W^*=1$ and 8 to mimic prolate and oblate HBPs, respectively, and added $W^*=\sqrt{L^*} \approx 3.46$ for self-dual shaped HBPs, a geometry exactly in between the oblate and prolate shape. The so-defined nine systems were all in the $\rm I$ phase. In the $\rm I$ mixtures of Fig.\,\ref{fig06:MSDI_c}, deviations from Gaussian dynamics, where $\alpha_{2,3} > 0$, become significant at $t/\tau>1$ for $W^*=1$ and $t/\tau>10$ for $W^*=8$. To explore the same time scales in pure systems of HBPs, we set the MC time step to $\delta t_{MC,c} = 1.0 \times 10^{-2}$. The resulting MSDs and NGPs are shown, respectively, in the top and bottom frames of Fig.~\ref{fig08:MSD_CL}. As expected, the MSDs corresponding to the same particle geometry collapse on a single curve at short times, when the effect of packing is still negligible. As time increases, a moderate deviation from the linear regime is observed at all concentrations, with the long-time MSD decreasing with increasing system packing. This decrease in MSD with packing correlates very well with the increase in the peak of the NGP. 
While prolate HBPs exhibit a relatively pronounced non-Gaussianity at $\eta=0.15$ and 0.20 (see frame (d) in Fig.\,\ref{fig08:MSD_CL}), the NGP of self-dual and oblate HBPs is less significant at all  packing fractions, suggesting a quasi-Gaussian dynamics. 

Typically, large values of the NGP at intermediate time scales stem from the occurrence of transient cages. If the morphology of these cages followed a common pattern, independent from their location in the system, deviations from a Gaussian behaviour would be very limited, even at large system densities. In $\rm N_U$ phases, HBPs are almost perfectly aligned as the large values of their order parameters confirms \cite{Cuetos2017}. This  suggests that cages are very similar to each other, being the random orientation of the particle minor axes possibly the only discerning feature. By contrast, in $\rm I$ phases, particles are randomly oriented and, if the resulting NGP is large, such a structural randomness must be somehow transferred to their dynamics. In $\rm I$ phases of 4-n-hexyl-4'-cyanobiphenyl, a prolate molecule able to form LCs, the occurrence of anomalous diffusion was ascribed to the presence of cages comprising molecules that tend to align in the direction of their longitudinal axis \cite{DeGaetani2007}. If this picture holds at the colloidal scale, one might expect the occurrence of clusters of similarly oriented HBPs with broad enough size and shape distributions to spark non-Gaussian dynamics. To test these intuitions, we have verified whether clusters actually exist in the $\rm I$ phase. To this end, following past works on the nucleation of crystals and nematic LCs in systems of colloidal rods \cite{Cuetos2007, Dijkstra2008}, we defined a criterion able to identify nematic-like clusters in $\rm I$ phases. According to this criterion, two HBPs $i$ and $j$ belong to the same cluster if their relevant axes ($\vers{e}_{L}$ for prolate and $\vers{e}_{T}$ for oblate HBPs) are sufficiently aligned. For prolate HBPs, $| \vers{e}_{L,i} \cdot \vers{e}_{L,j} | > K_1$. Additionally, the resulting order parameter of this cluster should be larger than a threshold value, i.e. $U_{L,local} > K_2$. Finally, $i$ and $j$ should be close enough along the cluster's nematic director, $\vers{n}_{local}$, and perpendicularly to it, i.e. $\| \vect{r}^{\parallel}_{i,j} \| < K_{\parallel,3}$ and $\| \vect{r}^{\perp}_{i,j} \| < K_{\perp,3}$. The threshold values, $K_1$, $K_2$, $K_{\parallel,3}$ and $K_{\perp,3}$, were first optimised to identify a cluster containing at least $95\%$ of HBPs in an $\rm N_U$ phase. The parameters optimised with this procedure are listed in Tab.\,\ref{tab07:opt_pars}, for both oblate and prolate HBPs.

\ctable[
 label = tab07:opt_pars,
 pos = ht!,
 caption = {Optimised parameters for nematic-like cluster identification in I phases. The parameterisation procedure has been performed in $\rm N_U$ phases of prolate and oblate HBPs at $\eta = 0.34$ (see text for details).},
 width = \columnwidth,
]{KKK}{
}{
                                      \FL
        $\text{Parameters}$   & $W^* = 1$ & $W^* = 8$ \ML
 $K_1$             & 0.85      & 0.85  \NN
 $K_2$             & 0.90      & 0.90  \NN
 $K_{\parallel,3}$ & 12.25     & 1.32  \NN
 $K_{\perp,3}$         & 1.55      & 13.00 \LL
}

\begin{figure}[!ht]
    \centering
    \includegraphics[width=\columnwidth]{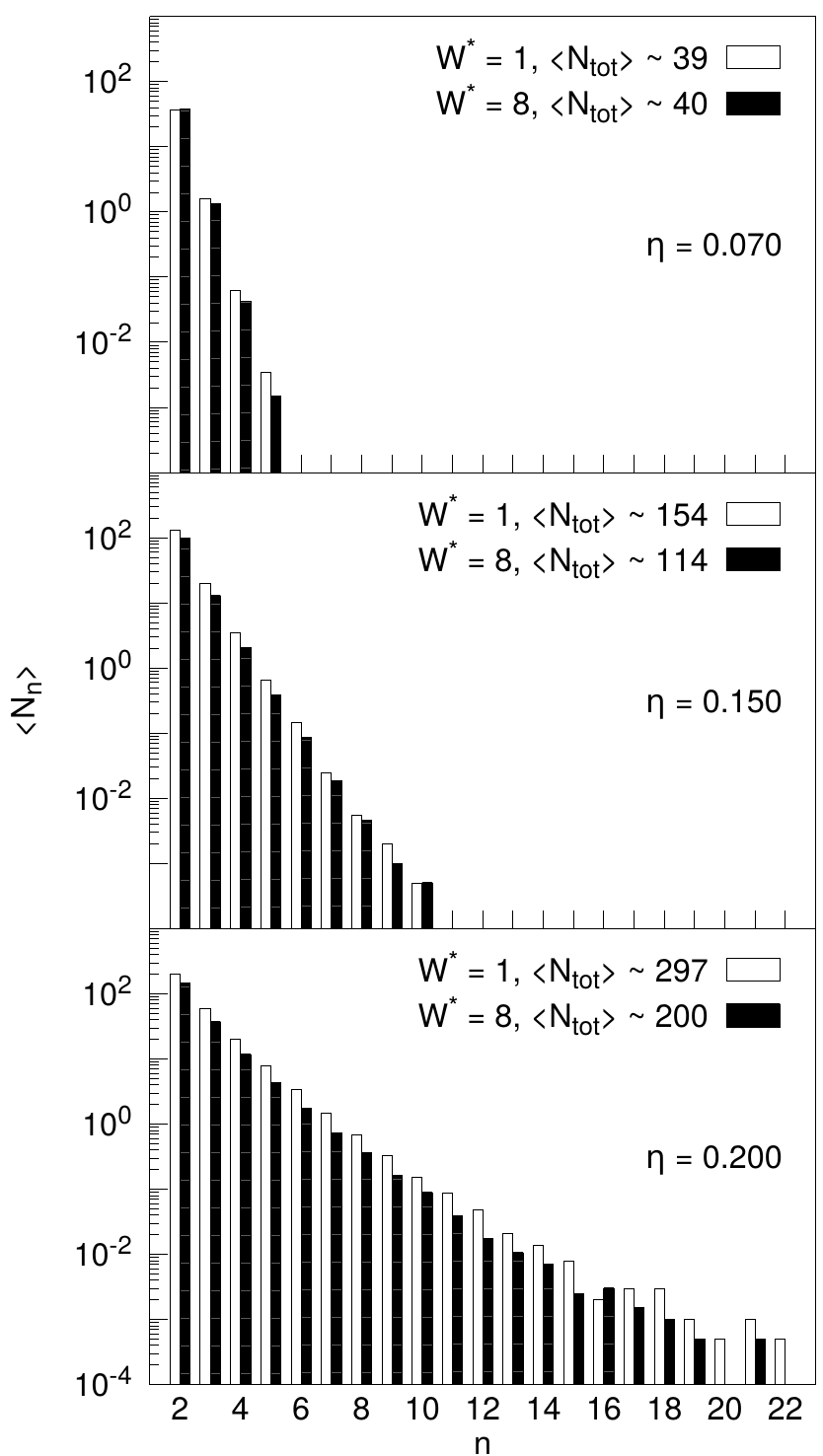}
    \caption{Average number of clusters $\langle N_n \rangle$ containing $n$ HBPs with $W^* =1$ (empty bars) and $W^* = 8$ (solid bars). All particle systems are made of $N = 2000$ HBPs forming $\rm I$ phases at packing fractions $\eta = 0.07$ (top frame), 0.15 (middle frame), and 0.20 (bottom frame). In the panel legend is reported the average number of clusters found in a sample configuration, i.e. $\langle N_{tot} \rangle$.}
    \label{fig09:CL_HIST}
\end{figure}

\begin{figure*}[!ht]
    \begin{subfigure}[b]{.3333\textwidth}
        \centering
        \includegraphics[width=\linewidth]{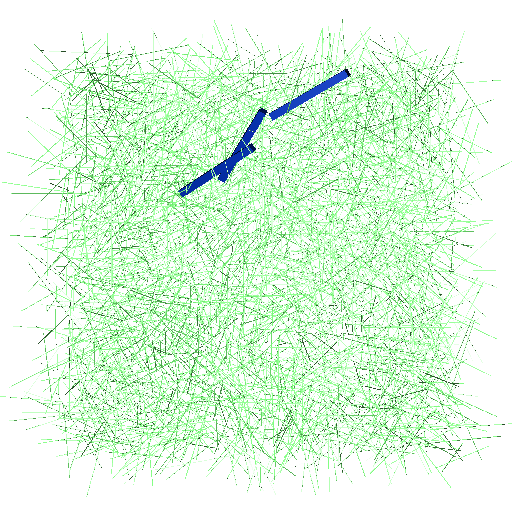}
        \caption{}
    \end{subfigure}%
    \begin{subfigure}[b]{.3334\textwidth}
        \centering
        \includegraphics[width=\linewidth]{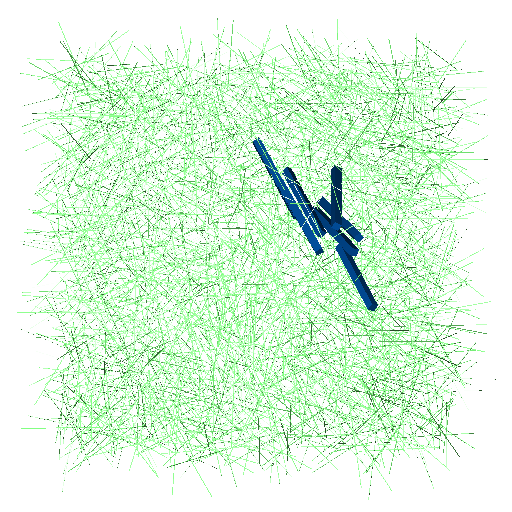}
        \caption{}
    \end{subfigure}%
    \begin{subfigure}[b]{.3333\textwidth}
        \centering
        \includegraphics[width=\linewidth]{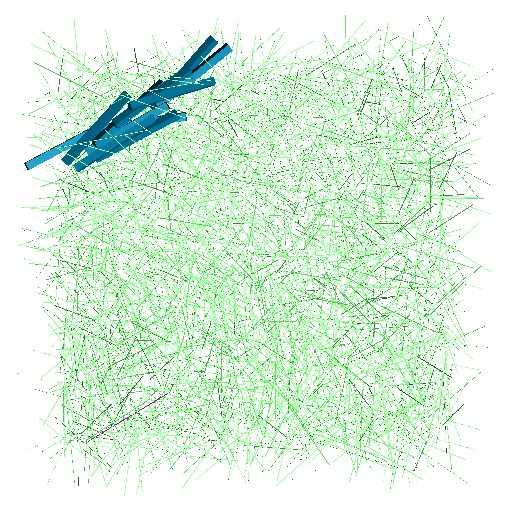}
        \caption{}
    \end{subfigure}
    \caption{Snapshots of HBPs with $W^* = 1$ in $\rm I$ phase with packing fraction $\eta = 0.20$. While clusters of sizes 3 (a), 7 (b) and 11 (c) are highlighted in blue, the remaining particles are represented as green lines.}
    \label{fig10:snap_cl_pro}
\end{figure*}

\noindent We then applied this criterion in $\rm I$ phases of prolate and oblate HBPs to ponder the existence of nematic-like clusters and estimate their size distribution. Results are shown in Fig.\,\ref{fig09:CL_HIST} for both prolate and oblate HBPs at the three packing fractions studied. Most clusters contain no more than $n=5$ particles at the largest system density and this number reduces to $n=2$ in very dilute $\rm I$ phases. While at low particle concentrations the average cluster size, $\langle N_n \rangle$, decays rapidly with $n$, the cluster size distribution becomes broader as the particle concentration increases. Consequently, as system packing increases, more and larger clusters are observed. Cluster distributions of prolate and oblate systems are very similar and, excluding the more dilute suspension at $\eta=0.07$, prolate HBPs are slightly more prone to form clusters as the density increases. Differences in cluster formation can be better appreciated from the average total number of clusters, i.e. $\langle N_{tot} \rangle$, which reveals a larger number of clusters formed in systems of prolate HBPs at sufficiently large packing. To assess whether the orientation of the clusters is isotropic, we estimated the orientational distributions of the local nematic director along the Cartesian coordinates $\vers{x}, \vers{y}, \vers{z}$. Our results revealed that clusters are randomly oriented, with a uniform distribution of their directors (see Fig.\,S6 in Supplementary Material). Illustrative examples of nematic-like clusters forming in $\rm I$ phases are reported in Fig.\,\ref{fig10:snap_cl_pro} for prolate HBPs at $\eta = 0.20$ and in Fig.\,S7 for oblate HBPs at $\eta = 0.20$. The structure of the clusters in the snapshots suggest that the shape and size of the HBPs affect also the morphology of the clusters: prolate particles form prolate-like clusters, while oblate particles seem to arrange in oblate-structured clusters. Additionally, we observe that the presence of spherical tracers does not seem to play a significant role in the formation of clusters. This is evinced from the monotonic behaviour of many of the properties analysed with respect to density, together with uniform distributions of HSs observed in Fig.\,2. However, we stress that we have only investigated this effect at $\eta =0.150$, while at the remaining packing fractions the formation of clusters was analysed in pure systems of HBPs. A more thorough investigation of the dynamics of the clusters in the presence of spheres is necessary to address their stability over time in I phases.

\section{Conclusions}

In summary, we have analysed the dynamics of globular macromolecules, modeled as hard spherical tracers, in isotropic and nematic suspensions of colloidal HBPs. To this end, we have implemented DMC simulations, which allow one to replicate the Brownian motion of particles and cover a broad range of timescales. The small size of HSs compared to HBPs does not affect the structural and dynamical properties of the host phase, at least at the concentrations considered in this work. The MSD of HBPs and the corresponding apparent exponents, which measure the deviations from Fickian diffusion, show three distinct regimes: a linear behaviour of the MSD with time at short and long timescales, and a nonlinear trend at intermediate times representing sub-diffusive dynamics. This sub-diffusive behaviour develops simultaneously to the onset of deviations from Gaussian dynamics. The MSD parallel to the nematic director exhibits a more pronounced sub-diffusive region in systems of oblate HBPs. An opposite tendency is detected in systems of prolate HBPs. Consequently, prolate HBPs diffuse faster along $\vers n$, whereas oblate HBPs diffuse faster in the direction perpendicular to $\vers n$. This overall behaviour dominates the diffusion of guest HSs. On the one hand, while the NGPs of HSs perpendicular to $\vers n$ vanish with increasing particle width, parallel NGPs increase. On the other hand, the tracers qualitatively replicate the changes in diffusion coefficients with the geometry of the host HBPs. As $W^*$ increases, the diffusivities of HSs parallel and perpendicular to the $\vers n$ decrease and increase, respectively. This indicates that the diffusion of tracers depends on the structural organization of the systems in which they displace, and anisotropic suspensions of surrounding particles make the diffusion of the tracers anisotropic, with the formation of temporary cages only along specific directions. 

The packing fraction, the size and the shape of the HBPs guide the suspension to the formation of preferential orientations and structures, which enable preferential pathways for the tracers' diffusion. The s-VHFs of the tracers deviate slightly from the theoretical Gaussian distribution only at the same time where the correspondent $\rm NGPs>0$ and reach local maxima. The comparison of the total diffusion of HSs between $\rm I$ and $\rm N_U$ phases showed that larger system densities hamper the tracers's diffusion at long timescales, similar to experimental results on the dynamics in crowded suspensions of spherical particles \cite{Laurati2016}. The occurrence of cage effects in crowded suspensions is related to the percolation of the systems \cite{Sikorski2019}, which depends not only on the system density, but also on the shape of the colloidal particles and, as a consequence, on their structural organisation. In systems of prolate HBPs, we observed larger deviations of the total NGP in $\rm I$ phases rather than in $\rm N_U$ phases, even though I suspensions are less packed. By further investigating the possible origin of this interesting behaviour, we noticed that, in dense $\rm I$ phases, HBPs form nematic-like randomly-oriented clusters, whose size and morphology depends on particle anisotropy. We believe that the presence of these clusters determine significant deviations from Gaussian dynamics.


\section{Acknowledgements}
The authors acknowledge the Leverhulme Trust Research Project Grant RPG-2018-415 and the use of Computational Shared Facility at the University of Manchester.


\bibliography{REFERENCES}





\end{document}